\documentclass[%
 reprint,
 superscriptaddress,
showpacs,
 amsmath,amssymb,
 aps,
 pra,
]{revtex4-1}
\usepackage{graphicx}% Include figure files
\usepackage{dcolumn}% Align table columns on decimal point
\usepackage{bm}% bold math
\usepackage{color}
\usepackage[usenames,dvipsnames]{xcolor}
\usepackage[colorlinks = true, linkcolor = blue, citecolor = blue, urlcolor = blue]{hyperref}% add hypertext capabilities
\begin{document}

\preprint{APS/123-QED}
\title{Controlling the ellipticity of attosecond pulses produced by laser irradiation of overdense plasmas}

\author{M.~Blanco}
\affiliation{Photonics4Life reseach group, Departamento de F\'{i}sica Aplicada, Facultade de F\'{i}sica, Campus Vida, Universidade de Santiago de Compostela, 15782 Santiago de Compostela, Spain}
\author{M.T.~Flores-Arias}
\affiliation{Photonics4Life reseach group, Departamento de F\'{i}sica Aplicada, Facultade de F\'{i}sica, Campus Vida, Universidade de Santiago de Compostela, 15782 Santiago de Compostela, Spain}
\author{A.~Gonoskov}
\affiliation{Department of Physics, Chalmers University of Technology, SE-41296 Gothenburg, Sweden}
\affiliation{Institute of Applied Physics, Russian Academy of Sciences, Nizhny Novgorod 603950, Russia}
\affiliation{Lobachevsky State University of Nizhny Novgorod, Nizhny Novgorod 603950, Russia}

\date{\today}

\begin{abstract}

The interaction of high-intensity laser pulses and solid targets provides a promising way to create compact, tunable and bright XUV attosecond sources that can become a unique tool for a variety of applications. However, it is important to control the polarization state of this XUV radiation, and to do so in the most efficient regime of generation. Using the relativistic electronic spring (RES) model and particle-in-cell (PIC) simulations, we show that the polarization state of the generated attosecond pulses can be tuned in a wide range of parameters by adjusting the polarization and/or the angle of incidence of the laser radiation. In particular, we demonstrate the possibility of producing circularly polarized attosecond pulses in a wide variety of setups.

\end{abstract}

\maketitle

The generation of attosecond pulses during the ionization of noble gases by intense laser pulses \cite{Atto1, Atto2} has opened up wide opportunities for studying matter at previously unreachable attosecond time scales based on pump-probe metrology \cite{krausz.nrev.2014}. In these studies the attosecond pulses are mostly used as probes. The interaction of relativistically strong laser radiation with overdense plasmas provides a promising alternative for producing attosecond pulses of sufficiently high intensity that they can also be used as pumps. This can provide a principle tool for XUV-pump-XUV-probe metrology \cite{tzallas.natp.2011}. Controlling the polarization state of the generated attosecond pulses is interesting for these and other applications \cite{xu.natp.2012, gierz.nlett.2012, schutz.psc.1993, fan.pnas.2015, liu.prl.2011}. Such control has been recently demonstrated in simulations using overdense plasmas, which showed that by varying the interaction parameters it is possible to tune the polarization of the harmonics from linear to almost circular \cite{chen.ncom.2016, ma.oe.2016}. Similar results have been also obtained for gas targets \cite{CircularGas1, CircularGas2, CircularGas3}. However, in the case of overdense targets, the relatively high values of plasma density considered in these studies correspond to the regime of the relativistic oscillating mirror (ROM) \cite{bulanov.pop.1994, lichters.pop.1996, teubner.rmp.2009}, which is significantly less efficient than the regime of the relativistic electronic spring (RES) \cite{gonoskov.pre.2011}. The RES regime does not assume the limit of high density or the steep density profile that the ROM regime does, and therefore can be accessed with a realistic pre-plasma induced by finite laser contrast \cite{blackburn.arxiv.2017}. 

In this paper, we demonstrate how to control the ellipticity of attosecond pulses in a wide range of parameters related to the RES regime, in particular for the production of circularly polarized attosecond pulses. Moreover, we show that RES theory \cite{gonoskov.pop.2018} perfectly describes the polarization properties of the generated attosecond pulse train for arbitrary polarization and angle of incidence of the laser pulse. By choosing the interaction parameters accordingly, one can therefore produce XUV pulses of any preassigned ellipticity.

High-order harmonic generation (HHG) from the surface of an overdense plasma irradiated by a relativistically strong laser pulse is a nonlinear phenomenon that is well-known for several decades \cite{HHG_old_1, HHG_old_2, HHG_old_3}. It has been especially well analysed under the assumption of steep density profiles and high plasma densities characterized by $S \gtrsim 10$. Here $S = n/a$ is the similarity parameter, with $n$ the plasma density in critical density units and $a$ the radiation amplitude in relativistic units. In this case, the interaction can be well described by the ROM model \cite{bulanov.pop.1994, lichters.pop.1996, teubner.rmp.2009, baeva.pre.2006, pirozhkov.pop.2006, debayle.pop.2013, boyd.pla.2016}, which phenomenologically assumes equating the electromagnetic fluxes of the incoming and outgoing signals at some point oscillating around the plasma surface. Apart from its theoretical study, over the last decade the generation of high harmonics in laser-plasma interaction has been a matter of intense experimental studies \cite{experimental_1, experimental_2, experimental_3, experimental_4, experimental_5}.

With decrease of the $S$ parameter (i.e. increase of intensity and/or decrease of density) from $S \sim 10$ to $S \sim 1$, the nature of plasma dynamics changes, becoming radically different from that implied by the ROM model. Instead of repelling energy instantaneously, within each cycle the plasma first accumulates up to 60\% of the incident energy and only then re-emits it back, acting as a spring \cite{gonoskov.pre.2011}. This can happen in the interaction of a solid-density plasma with a laser pulse that has finite contrast: the electron density can satisfy $S \sim 1$ locally, as a hot preplasma expands into vacuum \cite{blackburn.arxiv.2017}. The temporarily accumulated energy is stored in the form of electromagnetic fields, which arise when the radiation grabs and shifts inward frontier electrons relative to less mobile ions. In these situations, the shifted electrons tend to form a thin sheet, which moves so that its emission compensates the incident radiation in the plasma bulk. It is this phenomenological principle that underlies the RES theory \cite{gonoskov.pre.2011, gonoskov.pop.2018}. As the RES based theoretical model describes and provides insight into the variety of interaction scenarios, this regime can be referred to as the RES regime. A notable prediction is the generation of attosecond bursts with amplitude several times higher than the incident radiation \cite{gonoskov.pre.2011, fuchs.epjst.2014, bashinov.epjst.2014}. The generation mechanism includes three stages: (1) energy accumulation due to the shift of the frontier electrons, (2) transferring the accumulated and incident energy to the thin sheet of electrons during its backward motion and (3) emission of the burst by the sheet driven and energetically fed in this way. The last stage is also referred as coherent synchrotron emission (CSE) \cite{anderbrugge.pop.2010} and has also been studied in the case of laser interaction with a thin foil \cite{mikhailova.prl.2012, bulanov.pop.2013}.

Although the RES principle is valid for arbitrary incident polarization, all previous studies of this process have been done for the case of oblique incidence of P-polarized laser radiation, which naturally leads to linear polarization of the generated attosecond bursts. To analyze the effect of polarization we consider the problem using a rectangular coordinate system in a reference frame \cite{bourdier.physfluids.1983} moving along the plasma surface so that instead of oblique incidence we have normal incidence (towards the positive $x$ direction) onto plasma streaming towards the negative $y$ direction.

In the case of P-polarization, the electric field vector of the incident radiation is oriented along the $y$ axis just as it is for the radiation that corresponds to the flow of uncompensated ions (in the region from which the electrons have been evacuated). Thus, in order to compensate the radiation in the plasma bulk the electrons in the sheet, apart from shifting in the $x$ direction, should move only along the $y$ direction, producing radiation with only non-zero $y$ component of the electric field towards both positive and negative $x$. The former provides the compensation, while the latter appears as the generated linearly polarized attosecond bursts. Note that in this case the Lorentz force does not accelerate electrons in $z$ direction (this kind of consistency of the RES principle with first principles has also been shown in \cite{serebryakov.pop.2015}).

In the case of arbitrary polarization, to compensate the $z$ component of the electric field of the incident radiation, the sheet has to move in the $z$ direction as well. If both the $y$ and $z$ components of the sheet's velocity pass simultaneously through the vicinity of zero, the sheet moves with speed close to the speed of light in the negative $x$ direction, emitting a short burst with singularly strong $y$ and $z$ electric field components. The ellipticity of the high-harmonics in this process arises from the asymmetry of passing the vicinity of the zero point and depends on the interaction parameters in a complex way.

To study the generation of attosecond pulses with complex polarization states in the RES regime, we use the RES theory, accounting for the motion of the electrons in $z$ direction \cite{gonoskov.pop.2018}. Prior to presenting the results of parametric study we use \textit{ab initio} PIC simulations to demonstrate that (1) the RES theory accurately describes the overall plasma dynamics and (2) the results can be scaled to different plasma densities and laser amplitudes as they predominantly depend only on the parameter $S$.
	
\begin{figure}[htbp]
	\centering
	\includegraphics[width=\linewidth]{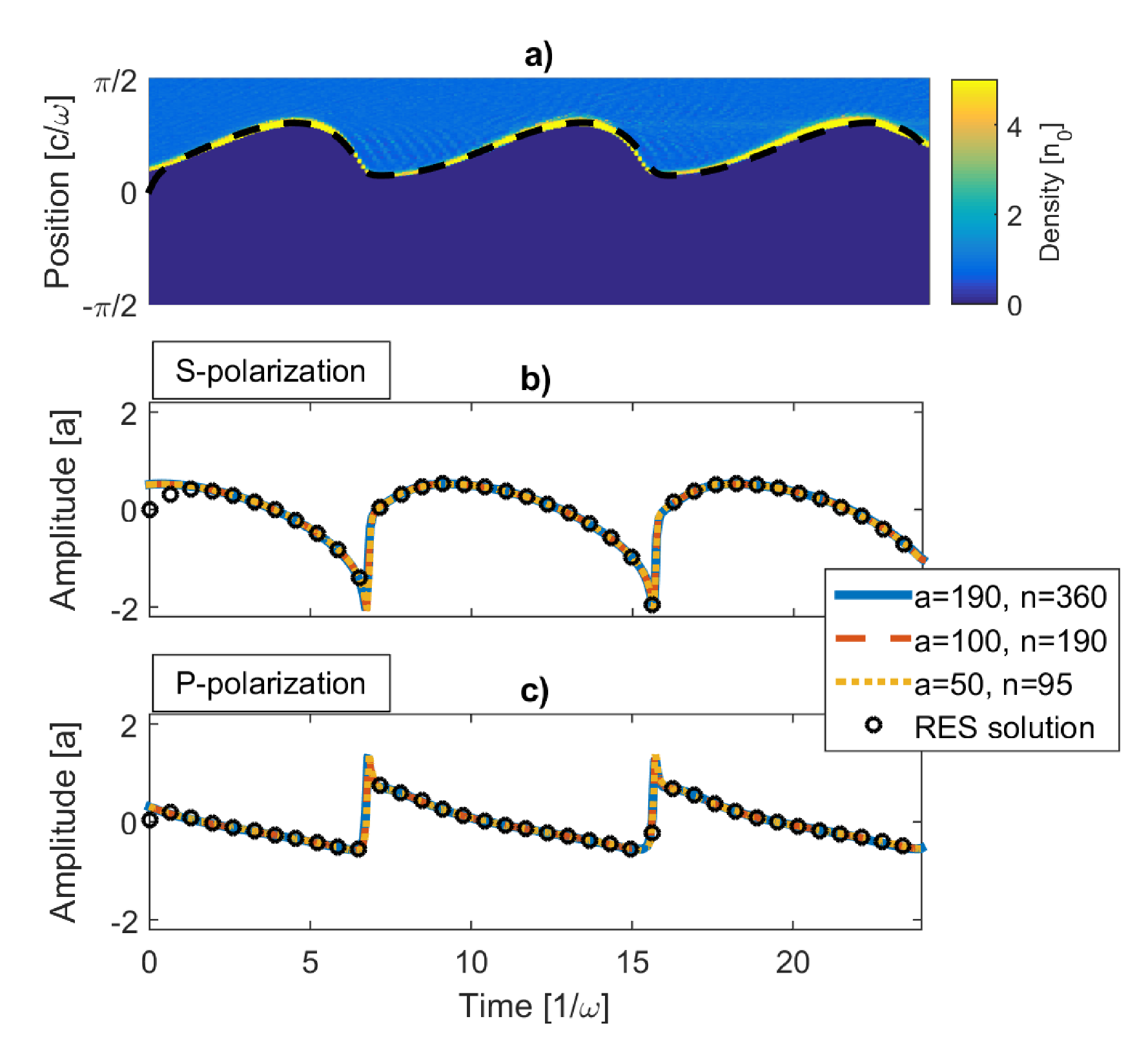}
	\caption{The results of PIC simulations for several amplitudes $a$ (see values in the insert) of a circularly polarized pulse incident obliquely onto a plasma with the proportional density $n$ so that $S = n/a = 1.9$: the electron density distribution in the boosted frame (a) and the reflected electric field components in S-polarization (b) and P-polarization (c). The results are in perfect agreement with the RES theory calculations that are shown through (a) the thin sheet position (dashed curve) and (b, c) the reflected field components (circles) as functions of time.}  
	\label{fig:fig1}
\end{figure}

For this purpose, we present three PIC simulations which have different values of radiation amplitude $a$ and plasma density $n$, but the same value of $S = n/a$. In the simulations the incident laser radiation is circularly polarized with a rectangular temporal profile and a duration of 3 periods. It has an amplitude $a$ of 190, 100 and 50 and is incident at an angle of $\theta = 45^\circ$ onto a plasma with a steep density profile and a density $n$ of 360, 190 and 95, respectively. Here $a$ is given in units of the relativistic amplitude $mc\omega/e$ and $n$ in units of the critical density $n_c = m \omega^2/\left(4 \pi e^2\right)$, where $\omega = 2\pi c/(0.8~\rm \mu m)$ is the laser frequency, $c$ is the speed of light, $m$ and $e$ are the mass and charge of the electron, respectively. PIC simulations were performed using the code PICADOR \cite{bastrakov.jcs.2012, surmin.cpc.2016}. To simulate oblique incidence with only one spatial dimension, we have used the boosted frame method, that consists in making a Lorentz transformation with a velocity of $c \sin\theta$ \cite{bourdier.physfluids.1983}. Figure \ref{fig:fig1} shows (a) the temporal evolution of the electron density and (b, c) the y- and z-components of the reflected/re-emitted electric field (hereafter referred to as the S- and P-polarized components, respectively). The simulation results are compared with the numerical solution of the differential equations of the RES theory (see Ref. \cite{gonoskov.pop.2018}) and perfect agreement can clearly be seen.

\begin{figure*}[hbtp]
	\centering
	\includegraphics[width=\linewidth]{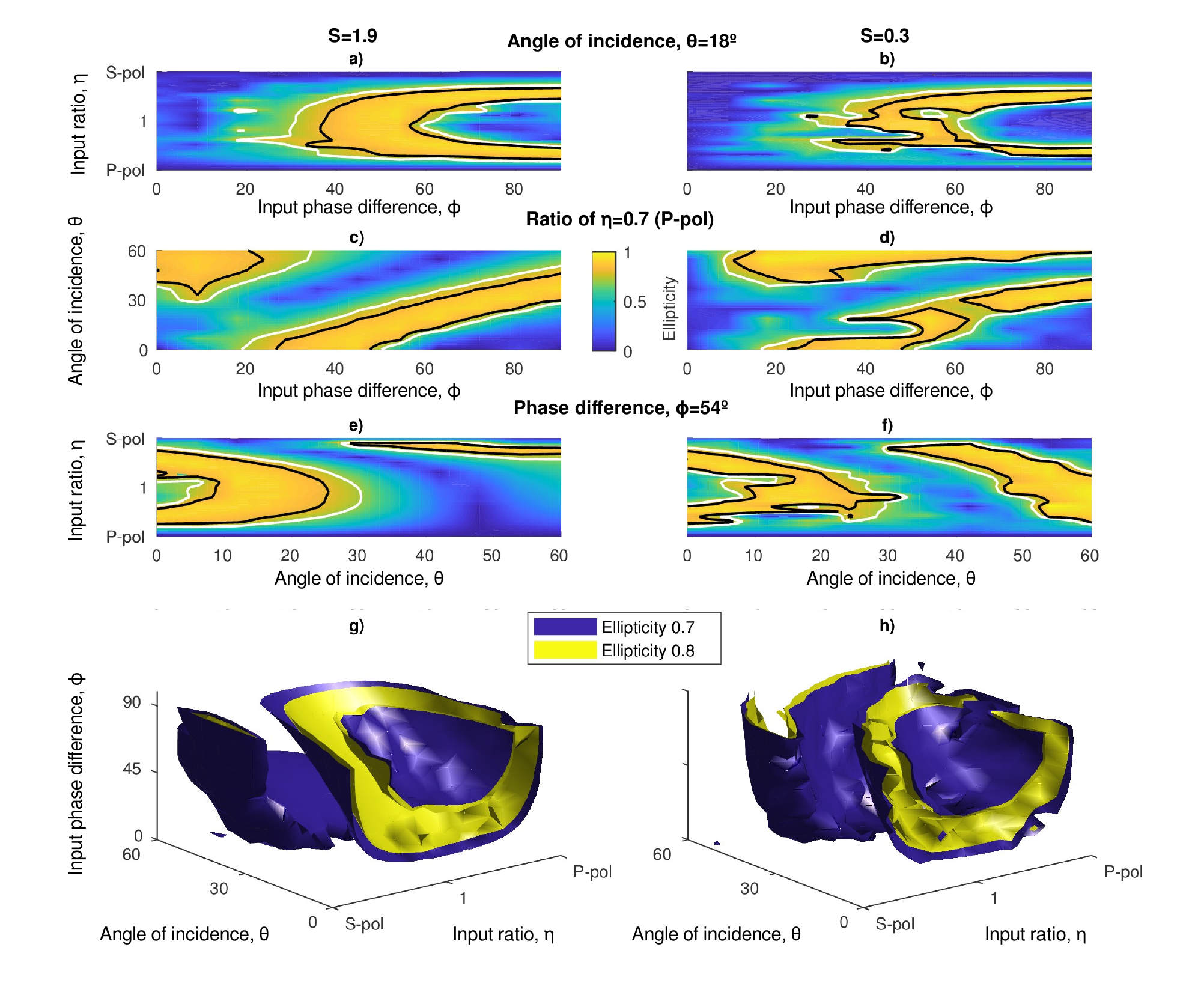}
	\caption{Ellipticity of the filtered attosecond pulses for $S = 1.9$ (a, c, e, g) and $S = 0.3$ (b, d, f, h). Results are shown for three examples of input parameters: (a, b) $\theta = 18^\circ$, (c, d) $\eta = 0.7$ with the P-polarized component being the major one and (e, f) $\phi = 54^\circ$. The contours indicate ellipticities of $0.7$ (white) and $0.8$ (black). The panels (g) and (h) show the isosurfaces for ellipticity equal to $0.7$ (blue) and $0.8$ (yellow) in the three-dimensional space of parameters.}
	\label{fig:fig2}
\end{figure*}

It is important to note that the different values of amplitude used in figure \ref{fig:fig1} can potentially affect the degree of ion motion and the role of radiation reaction. Both these effects are included in our PIC simulations. The three densities we have considered would arise from doubly, quadruply, and eight times ionized copper plasma. However, since we here consider only the physics of laser-plasma interaction, a full analysis of the realistic material density and the level of ionization are outside the scope of this paper. Figure \ref{fig:fig1} demonstrates that the results are self-similar at least down to $a = 50$ and are not altered significantly due to ion motion and radiation reaction for amplitudes at least up to $a = 190$.

Based on the obtained agreement we can now use the RES theory for assessing the opportunities for varying the ellipticity of the attosecond pulses through the variation of the interaction parameters. To characterize the polarization state of the laser radiation we use two parameters: (1) the ratio $\eta$ of the smaller of the P-polarized and S-polarized component amplitudes to the larger of the two and (2) the phase difference $\phi$ between the components. We vary the angle of incidence $\theta$, the ratio $\eta$, and the phase difference $\phi$. The parameter space spans angles of incidence $\theta$ between $[0^\circ,~60^\circ]$, phase differences $\phi$ between $[0^\circ,~90^\circ]$ and ratios for the component amplitudes between $[0,~1]$. We consider both the case when the amplitude of the P-polarized component is larger than that of the S-polarized component and the case where it is smaller. In the limit of $\eta = 0$, these cases correspond to P-polarization and S-polarization, which we use to denote them on the diagrams. For the outgoing radiation we filter out the harmonic orders outside the range of $\omega \in [30,~60]$ and calculate the ellipticity of the resulting waveform. These calculations are done for two values of similarity parameters $S = 1.9$ and $S=0.3$.

The results of the parametric scan are shown in figure \ref{fig:fig2}. The top panels of figure \ref{fig:fig2} show 2D maps of the ellipticity of the filtered attosecond pulses (in each row of the figure, a different parameter of the scan is held fixed at an arbitrarily chosen value). The results for the ellipticity in the three-dimensional space of parameters are shown in the last row of figure \ref{fig:fig2}, showing isosurfaces for the ellipticity of the attosecond pulses. We can distinguish two isolated prominent regions in parameter space that correspond to the generation of near-circularly polarized attosecond pulses. The first region spans from $\theta = 0$ to $\theta \approx 40^\circ$ for both $S = 0.3$ and $S = 1.9$. The boundary of this region can be roughly fitted by the expression

\begin{equation}
	\left(\frac{1 - \eta}{0.7}\right)^2 + \left(\frac{90 - \phi}{54}\right)^2 = \left(\frac{64 - \theta}{57}\right)^2
\end{equation}
 
\noindent which can be used to guide experiments and the development of a reliable source of circularly polarized XUV pulses.

The second region appears only for incidence angles $\theta \gtrsim 45^\circ$ and is broader for $S = 0.3$ than for $S = 1.9$. The fact that it is more prominent for small values of $S$ shows that this region appears as a unique feature of the RES regime. In contrast to the first region, the second region can span to $\phi = 0$ (see, for example, $\phi$-$\theta$ map for $S = 1.9$). This means that linearly polarized laser pulses under optimal orientation of the target can provide the generation of almost circularly polarized attosecond pulses.

\begin{figure*}[htbp]
	\centering
	\includegraphics[width=\linewidth]{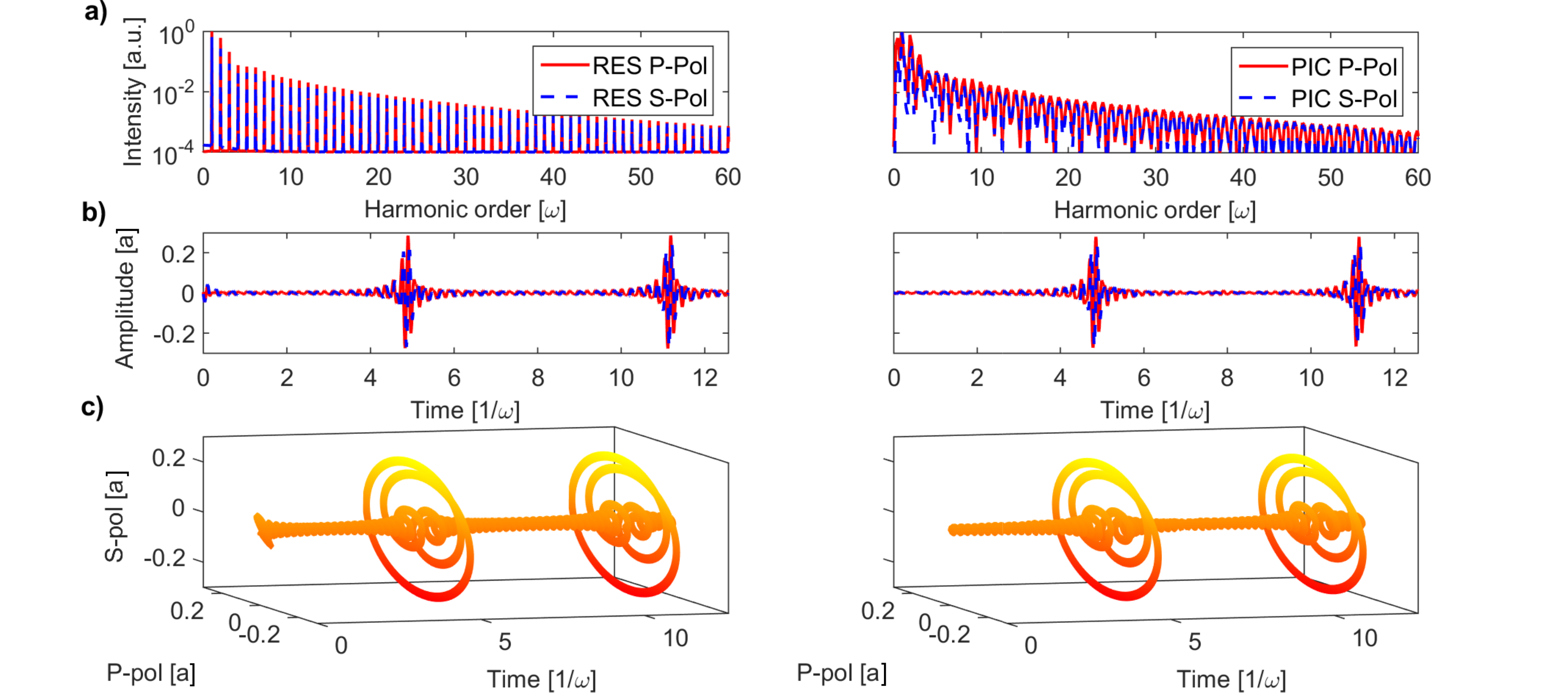}
	\caption{Circularly polarized pulses obtained for the optimal parameters when $S=1.9$. The left column represents the result from RES calculations and the right column those from PIC simulations. In each row it is shown (a) the reflected spectral intensity for both polarization components, normalized to the fundamental harmonic order, (b) the temporal shape of the pulses obtained from filtering the harmonic orders between $30$ and $60$ and (c) a 3D view of these pulses, to better highlight their polarization state.}
	\label{fig:fig4}
\end{figure*}

We have demonstrated that circularly polarized attosecond pulses can be produced in a variety of configurations. However, the emitted amplitude varies across these configurations. For $S = 1.9$ the highest amplitude is obtained for $\theta = 36^\circ$, $\phi = 81^\circ$ and $\eta = 0.5$, where the P-polarized component is larger than the S-polarized component. For these parameters the generated pulse has duration of $\tau=85$~as (FWHM for intensity), ellipticity of $\varepsilon=0.84$ and amplitude of $0.29a$, which is higher than reported elsewhere \cite{chen.ncom.2016, ma.oe.2016}. Figure \ref{fig:fig4} illustrates the pulse train for this case. It displays the spectral and temporal form of a pulse train, for the harmonic orders between 30 and 60, for both the RES model calculations and PIC simulations, showing good agreement.

We conclude that in the RES regime of laser-plasma interaction the ellipticity of generated attoseconds bursts can be tuned over a wide range of values by adjusting the target orientation and the ellipticity of the incident laser radiation. Achieving near-circular polarization is possible in a variety of configurations within two regions in parameter space: one region appears for incidence angles $\theta > 45^\circ$ and another for $\theta < 40^\circ$. The region for large angles becomes more prominent for small values of the relativistic similarity parameter $S$ and thus can be uniquely attributed to the RES regime. In particular, we note that production of attosecond pulses with near-circular polarization is possible with linearly polarized laser radiation if the target is oriented appropriately. The region for small angles does not alter significantly with change of $S$. We have provided a fit to the boundary of this region that can be used as universal robust guidance for experiments and future developments. Furthermore, we have demonstrated the utility of RES theory, as it is shown to agree perfectly with PIC simulations.

\section*{Acknowledgments}
This work has been partially funded by the Spanish Ministry of Economy and Competitivity (MINECO) under project MAT2015-71119-R, by Xunta de Galicia/FEDER under project ED431B 2017/64, by the Knut and Alice Wallenberg project PLIONA and by the Swedish Research Council Grants No. 2012-5644, No. 2013-4248 and No. 2017-05148. Manuel Blanco thanks the FPU grant program from the Spanish Ministry of Education, Culture and Sports (MECD). A.~G. thanks T.~G.~Blackburn for useful discussions.

% Bibliography
\bibliography{biblio1}{}
\bibliographystyle{apsrev4-1}

\end{document}